# A Metal-Insulator Transition via Wigner Crystallization in Boron Triangular Kagome Lattice

Woo Hyun Han[1], Sunghyun Kim[2], In-Ho Lee[3], and Kee Joo Chang[1,*]

[1]*Department of Physics, Korea Advanced Institute of Science and Technology, Daejeon 34141, Korea*

[2]*Department of Materials, Imperial College London, London SW7 2AZ, United Kingdom*

[3]*Korea Research Institute of Standards and Science, Daejeon 34113, Korea*

*e-mail: kjchang@kaist.ac.kr

**Abstract**

The flat band has attracted a lot of attention because it gives rise to many exotic phases, as recently demonstrated in magic angle twisted bilayer graphene. Here, based on first-principles calculations, we identify a metal-insulator transition in boron triangular Kagome lattice with a spin-polarized flat band at 2/3-filling. This phase transition is accompanied by the formation of a Wigner crystal, which is driven by Fermi surface nesting effect and thereby strong electron-phonon interactions, keeping ferromagnetism. Our calculation results suggest that boron triangular Kagome lattices with partially filled flat bands may open a new playground for many exotic quantum phases in two-dimensional systems, such as Winger crystallization and fractional quantum Hall states.

The flat band is the hallmark of exotic quantum states in solids, for example, flat-band ferromagnetism [1,2], high-temperature superconductivity [3−6], Wigner crystallization [7,8], and fractional quantum Hall (FQH) states [9−11], because electrons are confined in a narrow energy window with quenched kinetic energy and huge density of states. The flat bands mainly originate from destructive interference between hopping parameters in dice, Lieb, and Kagome lattices [12−14], but they also appear as non-trivial surface states, so-called drumhead-like surface states, in Dirac nodal line semimetals [15,16]. A recent theoretical study has shown that twisted bilayer graphene can possess flat bands at a magic twist angle [17], and subsequently, the accessibility of the flat bands and the electrical tunability through twist angle have been experimentally demonstrated [18−20]. This experiment has revealed the rich physics of partially filled flat bands, showing unconventional superconductivity as well as strongly correlated phases at half-filling [18−20].

A Kagome lattice is a prototype system with intensive studies on flat band physics [1, 9−10,21−32], which have suggested various phases. At half-filling of the flat band, it is known that ferromagnetic ordering is preferred in Kagome lattices, so-called flat-band ferromagnetism [1,25]. For electron densities lower than half-filling, for example, 1/6- and 1/9-filling, the formation of a Wigner crystal has been reported, in which interaction energies dominant over kinetic energies, leading to a new crystal phase [26,27], whereas magic angle twisted bilayer graphene (MA-TBG) shows a Mott-like insulating gap at half-filling [19]. Moreover, through first-principles calculations, quantum spin Hall and quantum anomalous Hall states for non-trivial flat bands have been suggested in Li-doped herbertsmithites [30], metal-organic frameworks [31], and boron triangular Kagome lattice [32]. In addition to exploring new phases, it is natural to ask how partial filling of the flat band affects the structural stability and induces exotic phase transitions in real materials when tuning the Fermi level within the flat band. However, due to the limited material realization of stable Kagome lattices, the dynamical stability and evolution of new phases as a function of flat band filling have been remained unclear yet.

In this Letter, through first-principles calculations, we investigate the electronic structure and dynamical stability of a two-dimensional (2D) boron triangular Kagome lattice with a partially filled flat band, which has been recently predicted [32]. Under hole doping into the spin-polarized flat band, we identify that the Fermi surface nesting effect becomes significant, yielding strong electron-phonon interactions. At 2/3-filling of the spin-polarized flat band, the dynamical instability occurs in a metallic ferromagnetic phase. However, instead of structural transformation, we find a metal-insulator transition accompanied by the formation of an insulating Wigner crystal phase. Our results suggest that the 2D boron triangular Kagome lattice offers a new perspective to explore new exotic phases by tuning the filling factor of the flat band.

For a normal $B_3$ Kagome lattice, which consists of corner-sharing triangles, a simple tight-binding model with only nearest-neighbor interactions exhibits a completely flat band and two dispersive bands that cross at the K points in the 2D Brillouin zone (BZ), as shown in Fig. 1(a). The single flat band arises from the destructive interference of hopping between three atomic sites per cell. The flat band becomes slightly dispersive as next-nearest-neighbor interactions ($t_{\text{NNN}}$) are allowed, with its bandwidth increasing with the strength of $t_{\text{NNN}}$. When the Fermi level is lowered from the top of the flat band, the Fermi surface of the flat band evolves from a small circle to a hexagon around the Γ point. Then, there exists a nesting vector ($\boldsymbol{q}$) which connects the hexagonal sides of the Fermi surface. From the nesting function ($\chi_{\boldsymbol{q}}$) defined as,

$$\chi_{\boldsymbol{q}} = \sum_k \delta(\varepsilon_k - \varepsilon_F)\delta(\varepsilon_{k+q} - \varepsilon_F), \qquad (1)$$

the Fermi surface nesting is found to be prominent at the K point when the flat band is 2/3-filled [33]. Near $\boldsymbol{q} = 0$, the nesting function just represents the density of states around the Γ point. Since electron-phonon interactions are significantly affected by the Fermi surface nesting, the Fermi surface nesting plays an important role in determining phase transitions such as charge-density waves [34−38] and conventional superconductivity [39].

Recently, *ab initio* evolutionary crystal structure search calculations showed that a 2D boron triangular Kagome lattice composed of triangles in triangles [Fig. 1(b)], termed $B_9$ *t*-Kagome, can be synthesized on a silver substrate [32]. While an isolated $B_9$ *t*-Kagome lattice is slightly twisted, it turns into an ideal $B_9$ *t*-Kagome lattice under tensile strain above 16.5%. The band structure of the $B_9$ *t*-Kagome lattice near the Fermi level is similar to that of the $B_3$ Kagome lattice, exhibiting a nearly flat band. The spin-polarized flat bands are well separated from each other, resulting in half-metallic ferromagnetism [32] [Fig. 1(b)], in contrast to MA-TBG that possesses a Mott-like half-filling insulating state [19]. Due to the existence of the nearly flat band, one can expect that hole doping will affect electron-phonon interactions and thereby induce a phase transition in the $B_9$ *t*-Kagome lattice.

To examine the effects of hole doping and tensile strain on the stability and magnetism of the $B_9$ *t*-Kagome lattice, we calculated electron-phonon coupling (EPC) matrices using a density functional perturbation theory for spin-polarized states. All density functional theory (DFT) calculations were performed by using the norm-conserving pseudopotentials and the Perdew-Burke-Ernzerhof functional for the exchange-correlation potential [40], as implemented in QUANTUM ESPRESSO package [41]. The total EPC constant ($\lambda$) was obtained by integrating the Eliashberg spectral function over frequency.

In experiments, the hole doping level ($n$ in units of hole per unit cell) can be controlled by varying an electrostatic gate voltage. As the hole doping level increases, i.e., the filling factor of the flat band ($\nu = 1 - n$) decreases, the density of states (DOS) at the Fermi level [$N(\varepsilon_F)$] rapidly increases, whereas the total magnetic moment ($\mu_{\text{tot}}$) decreases due to the decrease of majority-spin electrons [Fig. 1(c)]. In the $B_9$ *t*-Kagome lattice, tensile strain weakens the parameter $t_{\text{NNN}}$. Thus, as the tensile strain increases for a given doping level, the bandwidth of the flat band becomes narrower, resulting in the higher $N(\varepsilon_F)$. For excessive hole doping levels above the critical doping concentration, an imaginary phonon frequency occurs at the K point in the 2D BZ, indicating that the dynamical instability occurs (as will be discussed later). In the same manner, there exists a critical tensile strain for a given doping level. In this case, an imaginary phonon frequency appears at the Γ point, which results from the increase of the B-B bond lengths [33]. Within a moderate range of hole doping and tensile strain, we confirmed that the $B_9$ *t*-Kagome lattice maintains the ferromagnetism as well as the dynamical stability, as illustrated schematically in Fig. 1(d).

The calculated $\lambda$ values for various doping levels and different tensile strains are given in Table 1. For each tensile strain, there is a tendency that $\lambda$ rapidly increase with $n$. We find that the nesting function at the K point is significantly enhanced at a particular doping level of $n = 0.3$ hole/cell [Fig. 2(a)], while the nesting function near the Γ point grows monotonically. Figure 2(b) shows the distribution of the EPC constant in momentum space ($\lambda_q$). Its contributions from the Γ and K points are prominent, in good agreement with the general feature of the nesting function. The EPC constants were estimated to be 2.08, 3.48, and 8.10 for tensile strains of 18%, 20%, and 22%, respectively, at $n = 0.3$ hole/u.c. These values are comparable to those of other known materials such as atomic metallic hydrogen ($\lambda = 1.43 - 3.39$) [42], $MgB_2$ ($\lambda = 0.73$) [43], $H_3S$ at 200−250 GPa ($\lambda = 1.96 - 2.42$) [44], $YH_9$ at 150 GPa ($\lambda = 4.42$) [45], $YH_{10}$ at 400 GPa ($\lambda = 2.41$) [45], and 2D boron sheets ($\lambda = 0.37 - 1.05$) [46,47]. For MA-TBG with a flat band, recent theoretical calculations showed that the EPC constant is as high as 5, depending on the twist angle and carrier concentration [48]. In the $B_9$ *t*-Kagome lattice, the large enhancement of $\lambda$ by hole doping is attributed to the existence of the nearly flat band, similar to MA-TBG.

Figure 3(a) shows the phonon dispersion curves and phonon linewidths of the 18%-strained $B_9$ *t*-Kagome lattice at $n = 0.3$ and 0.4 hole/cell. We find that the localized $p_z$ orbitals of the flat band are coupled with two degenerate in-plane $E_{2g}$ modes denoted as *A* and *B* [33], similar to $MgB_2$, in which in-plane modes are coupled with σ-bonded electrons between the B atoms [43,49]. It is interesting to note that the mode *A* softens due to the Fermi surface nesting as $n$ exceeds 0.2 hole/cell [Fig. 3(b)].

Above the critical doping concentration of $n = 1/3$ hole/cell, corresponding to the filling factor $\nu = 2/3$, the imaginary phonon mode occurs at the K point, with the enhanced phonon linewidth. Although the large Fermi nesting function, which is an imaginary part of susceptibility, does not directly guarantee the structural instability, it is clear that the dynamical instability at the K point is caused by strong electron-phonon interactions [36,38].

When the imaginary phonon mode appears, the strained $B_9$ *t*-Kagome lattice cannot maintain the original translational symmetry. In order to search for other stable phases, we considered a larger 3×3 lateral supercell because the dynamical instability occurs at the K point. Above the critical doping level under 18% strain, we did not find any structural phase transition in the supercell calculations. Instead, we found a new magnetic phase with the local magnetic moments distributed asymmetrically. Since the primitive cell is a $\sqrt{3}\times\sqrt{3}$ supercell, as depicted in Fig. 3(c), three flat bands are formed due to the band folding effect [Fig. 4(a)]. Below the critical doping level, two flat bands are degenerate along the $M - K$ line, and the magnetic moments are equally distributed at the corner sharing B atoms forming an ideal $B_3$ Kagome lattice. In the new phase above the critical doping level, the twofold degeneracy of the flat band along the $M - K$ line is lifted by opening a gap, as shown in Fig. 4(a). At the filling factor just below $\nu = 2/3$, the Coulomb interaction causes the redistribution of electrons in the flat band, resulting in Wigner crystallization which breaks the original translational symmetry of the $B_9$ *t*-Kagome. The charge densities for the two occupied levels are localized with different weights at the B atoms forming an enlarged Kagome lattice and those of inner plaquettes along the hexagonal rings, which are marked by red and blue dots in the $\sqrt{3}\times\sqrt{3}$ supercell, respectively [Fig. 4(b)]. Recently, a Wigner crystal has been suggested for carbon Kagome graphene with the partially filled flat band [27]. Our Wigner crystal has close-packed patterns on a triangular lattice, similar to that in carbon Kagome lattice. In carbon Kagome lattice, an additional 1/3-filling for the spin-polarized flat band was required to achieve the Wigner crystallization after one-hole doping that leads to the half-metallic ferromagnetism [27]. In the $B_9$ *t*-Kagome lattice, the half-metallic ferromagnetic state already exists without doping, and the Wigner crystallization can be achieved below the critical filling of $\nu = 2/3$ for the spin-polarized flat band, as illustrated in Fig. 4(c). The enhanced electron-phonon interactions above $\nu = 2/3$ make the $B_9$ $t$-Kagome lattice unstable. However, the localized nature of the flat band favors the Wigner crystal without any structural transformation, in which the interaction energy dominates over the kinetic energy.

The band structures and spin configurations of the $B_9$ $t$-Kagome lattice with various filling factors are shown for the $\sqrt{3}\times\sqrt{3}$ supercell [Fig. 4(a)-(b)]. Above $\nu = 2/3$, the corner-sharing atoms have the same local magnetic moments, maintaining the original translational symmetry. On the other hand, below $\nu = 2/3$, where the Wigner crystal forms, the band gap opens near the Fermi level. The band gap tends to increase with decreasing of $\nu$, with the gap size of about 20 meV at $\nu = 0.6$. It is interesting to note that two bands lying just below the band gap cross each other at the K point, forming the Dirac point. In the Wigner crystal, the local magnetic moments in the enlarged Kagome lattice marked by red dots are larger than those of the inner hexagon plaquette marked by blue dots [Fig. 4(b)]. Although the total magnetic moment decreases with decreasing of $\nu$, the Wigner crystal maintains the ferromagnetic state. Due to the formation of an insulating Wigner gap, the hole doping actually induces a metal-insulator transition.

When a Kane-Mele type of spin-orbit coupling (SOC) [50] is included in the Wigner crystal phase, new quantum effects are expected due to the non-trivial band topology. When the filling factor is lowered to $\nu = 1/3$, the Fermi level can across the Dirac point at the K point. With including the SOC, we find that the band gap opens at the crossing points at the Γ and K points and three flat bands have the non-trivial Chern number of $C = 1$ [Fig. 4(a)] [33]. Although the SOC is small for B systems, its strength can be enhanced using various methods, such as hydrogenation, adatom deposition, and substrate proximity effects, which have been used for graphene [51, 52]. Since FQH phases can be observed for the non-trivial flat bands with $C = 1$ without magnetic fields [9-11], our calculations suggest that the Wigner crystal formed in the $B_9$ $t$-Kagome lattice can be a promising candidate for the FQH phase if a proper SOC is introduced.

In summary, through *ab initio* calculations, we have shown that the $B_9$ $t$-Kagome lattice with the spin-polarized flat band exhibits a metal-insulator transition from metallic ferromagnetism to an insulating Wigner crystal at 2/3-filling. While the dynamical instability occurs at 2/3-filling due to the large Fermi surface nesting and strong electron-phonon interactions, the formation of a Wigner crystal is preferred without any structural transformation. The Wigner crystal has a small charge gap with the asymmetric distribution of local magnetic moments in ferromagnetic ordering. Furthermore, the Wigner crystal has a potential to exhibit non-trivial topological flat bands in the presence of the spin-orbit coupling. A recent theoretical study has argued that the Mott-like insulating gap observed in half-filled MA-TBG may originate from Wigner crystallization, not from Mott insulation [53]. Our calculations suggest that examining the dynamical instability in MA-TBG as a function of carrier concentration may provide a clue to understanding the origin of the insulating gap at half-filling.

**Acknowledgments**

This work was supported by Samsung Science and Technology Foundation under Grant No. SSTFBA1401-08.

**Table 1.** Calculated electron-phonon coupling constants ($\lambda$) for various hole doping concentrations in $B_9$ *t*-Kagome lattice under tensile strain.

| Strain | 0 hole/cell | 0.1 hole/cell | 0.2 hole/cell | 0.3 hole/cell |
|---|---|---|---|---|
| 18% | 0.11 | 0.79 | 1.05 | 2.01 |
| 20% | 0.13 | 0.96 | 1.42 | 3.36 |
| 22% | 0.16 | 1.20 | 1.95 | 8.10 |

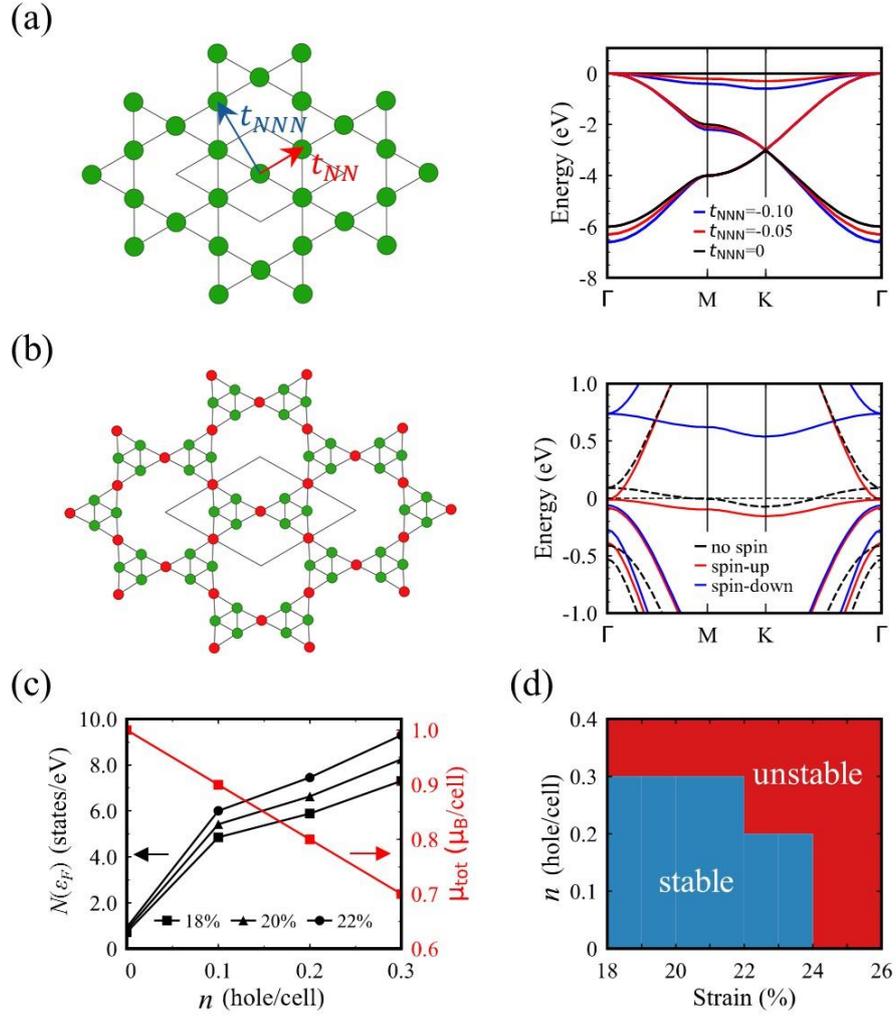

**Figure 1.** (a) The atomic structure of $B_3$ Kagome lattice and its tight-binding band structure for various values of $t_{NNN}$, with the nearest-neighbor hopping parameter $t_{NN} = -1$ eV. For $B_9$ $t$-Kagome lattice, (b) the atomic structure and the DFT band structure near the Fermi level, (c) the density of states at the Fermi level [$N(\varepsilon_F)$] and the total magnetic moment ($\mu_{tot}$) as a function of hole doping and strain, and (d) a schematic diagram for the dynamical stability of hole-doped $B_9$ $t$-Kagome lattice under strain. In (a) and (b), black parallelograms denote the 1×1 unit cell.

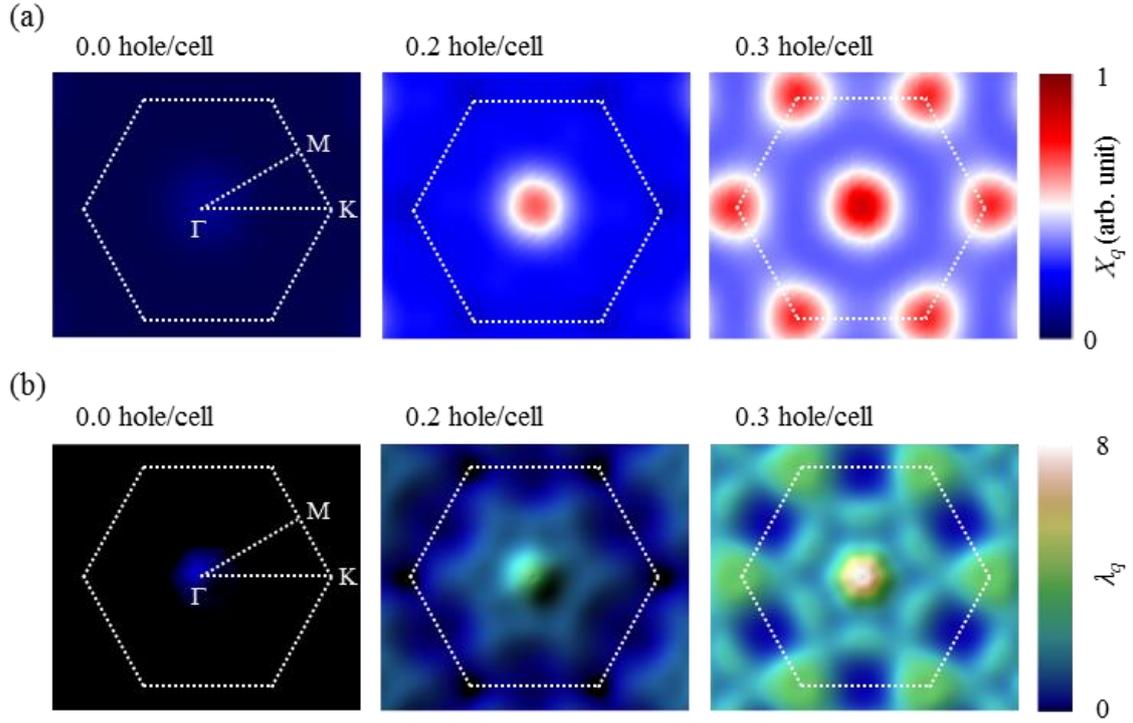

**Figure 2.** The distribution of (a) momentum-dependent nesting function ($\chi_{\boldsymbol{q}}$) and (b) electron-phonon coupling ($\lambda_{\boldsymbol{q}}$) in hole-doped B$_9$ *t*-Kagome lattices under 18% strain. The white dotted hexagons represent the 2D BZ.

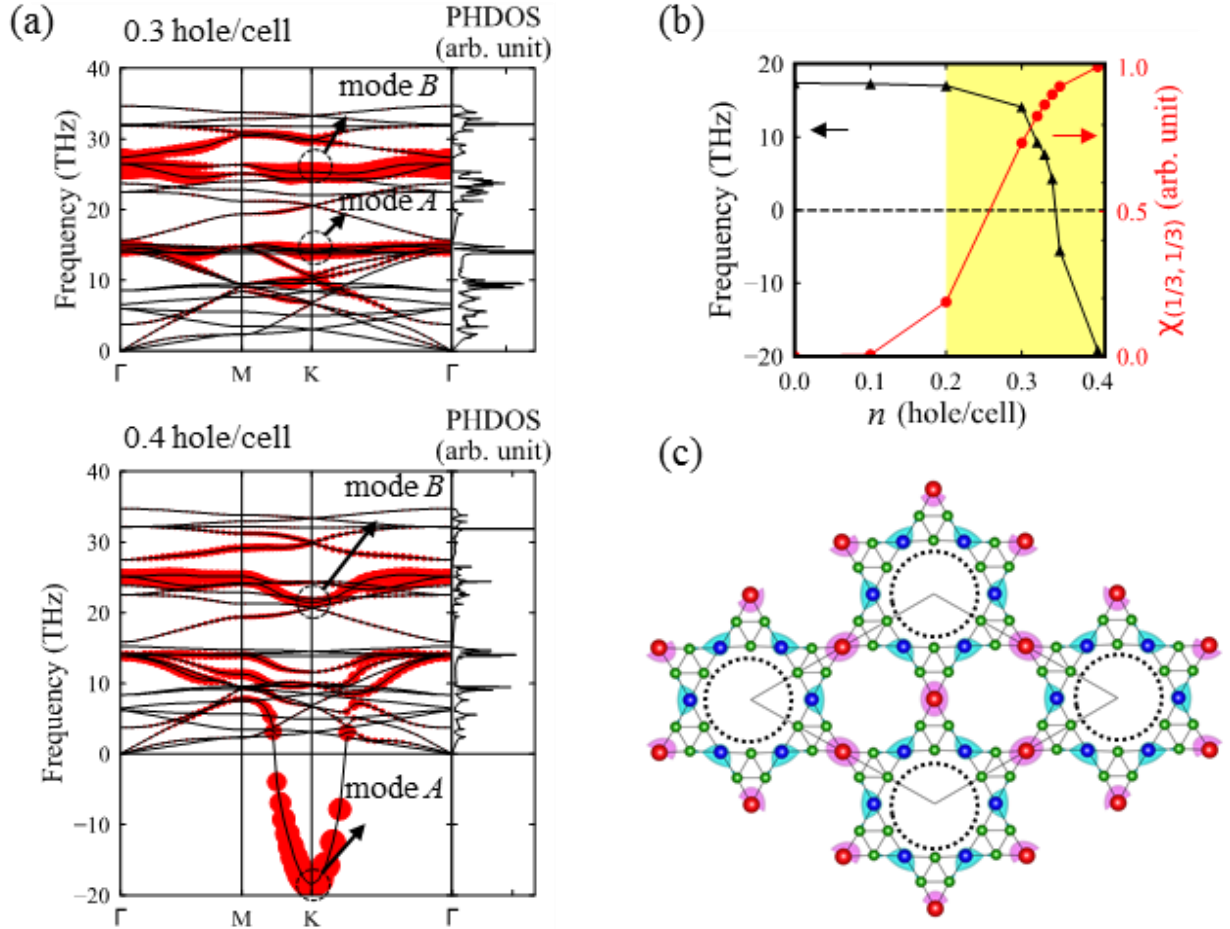

**Figure 3.** (a) The phonon dispersion curves and phonon densities of states (PHDOS) at hole carrier concentrations of 0.3 and 0.4 hole/cell in 18%-strained $B_9$ $t$-Kagome lattice. The red circles indicate the magnitude of phonon linewidths. (b) The variations of the nesting function and the phonon frequency of mode $A$ at the K point with hole carrier concentration. (c) The distribution of the charge densities for two occupied flat bands in the Wigner crystal at 2/3-filling. The red- and blue-colored circles represent the large and small local magnetic moments in Fig. 4(b), respectively. The black rhombus represents the $\sqrt{3}\times\sqrt{3}$ cell.

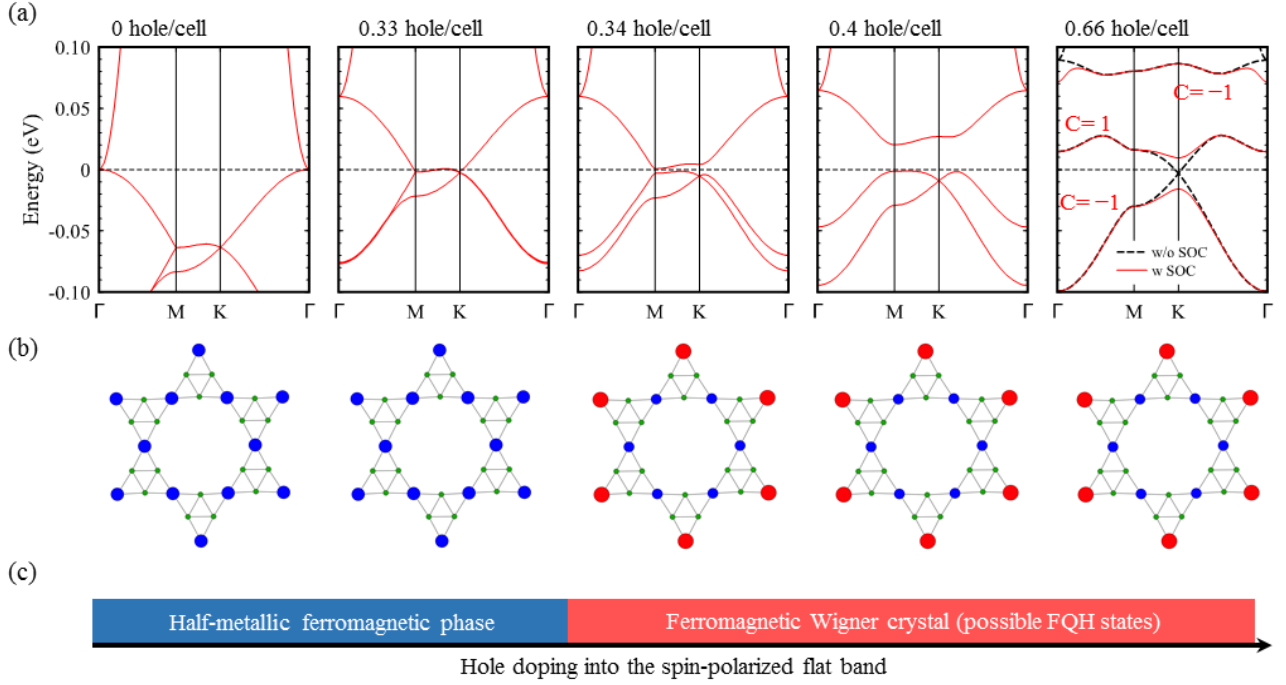

**Figure 4.** (a) The band structures of 18%-strained B$_9$ *t*-Kagome lattice in the 2D BZ of the $\sqrt{3}\times\sqrt{3}$ supercell for various hole carrier concentrations. (b) The distribution of local magnetic moments. The red- and blue-colored circles represent the large and small local magnetic moments, respectively, whereas the green-colored sites have zero magnetic moment. (c) A schematic diagram for the stable phases as a function of hole carrier concentration. The FQH states can be formed in the presence of the SOC.

# Supplementary Information

## A Metal-Insulator Transition via Wigner Crystallization in Boron Triangular Kagome Lattice

Woo Hyun Han[1], Sunghyun Kim[2], In-Ho Lee[3], and Kee Joo Chang[1,*]

[1]*Department of Physics, Korea Advanced Institute of Science and Technology, Daejeon 34141, Korea*

[2]*Department of Materials, Imperial College London, London SW7 2AZ, United Kingdom*

[3]*Korea Research Institute of Standards and Science, Daejeon 34113, Korea*

[*]Corresponding author. E-mail: kjchang@kaist.ac.kr. Telephone number: 82-42-350-2531. Fax number: 82-42-350-2510.

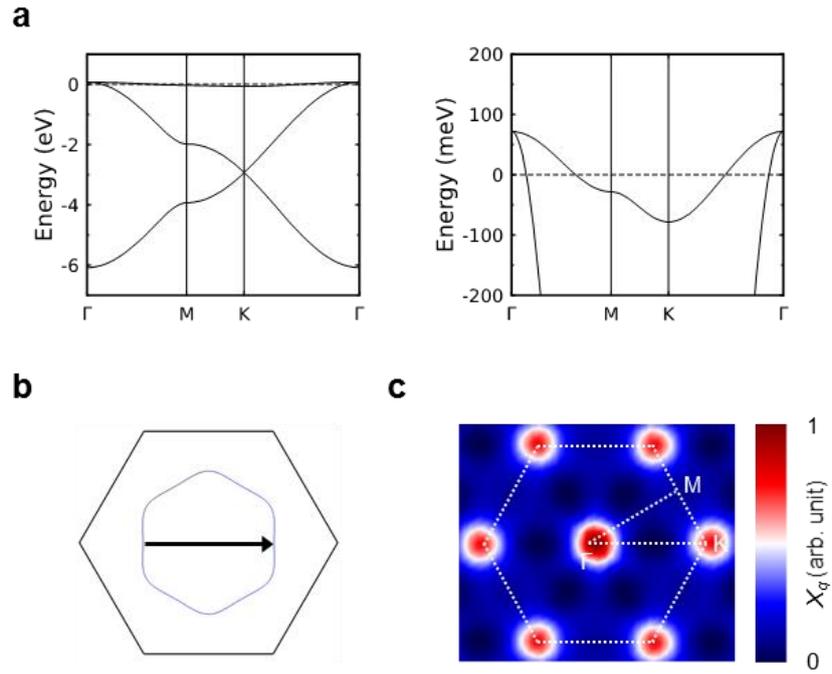

**Supplementary Figure 1.** (a) Tight-binding band structures, (b) Fermi surface, and (c) momentum-dependent nesting function ($\chi_q$) of $B_3$ Kagome lattice with the nearest-neighbor hopping parameter $t_{NN} = -1$ eV and the next-nearest-neighbor hopping parameter $t_{NNN} = -0.025$ eV, where the flat band is 2/3-filled.

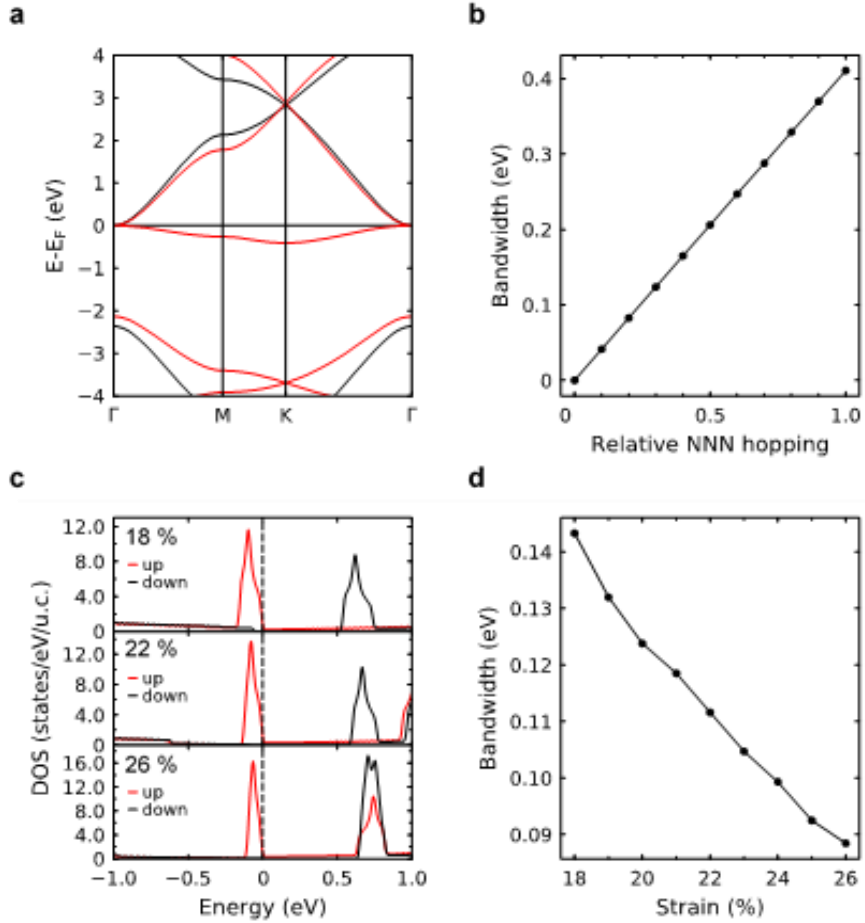

**Supplementary Figure 2.** Bandwidth as a function of next-nearest-neighbor hopping parameter and strain. For the $B_9$ *t*-Kagome lattice, (a) the tight-binding band structure using the nearest-neighbor (NN) and next-nearest-neighbor (NNN) hopping parameters obtained from the previous study [1] and (b) the bandwidth of the flat band as a function of the next-nearest-neighbor hopping parameter. In (a), the red and black solid lines indicate the band structures with and without the NNN hopping parameter, respectively. (c) The density of states (DOS) of the undoped $B_9$ *t*-Kagome lattice under various tensile strains through first-principles density functional calculations. (d) The bandwidth of the flat band near the Fermi level as a function of tensile strain. It is inferred that tensile strain decreases the strength of the NNN hopping.

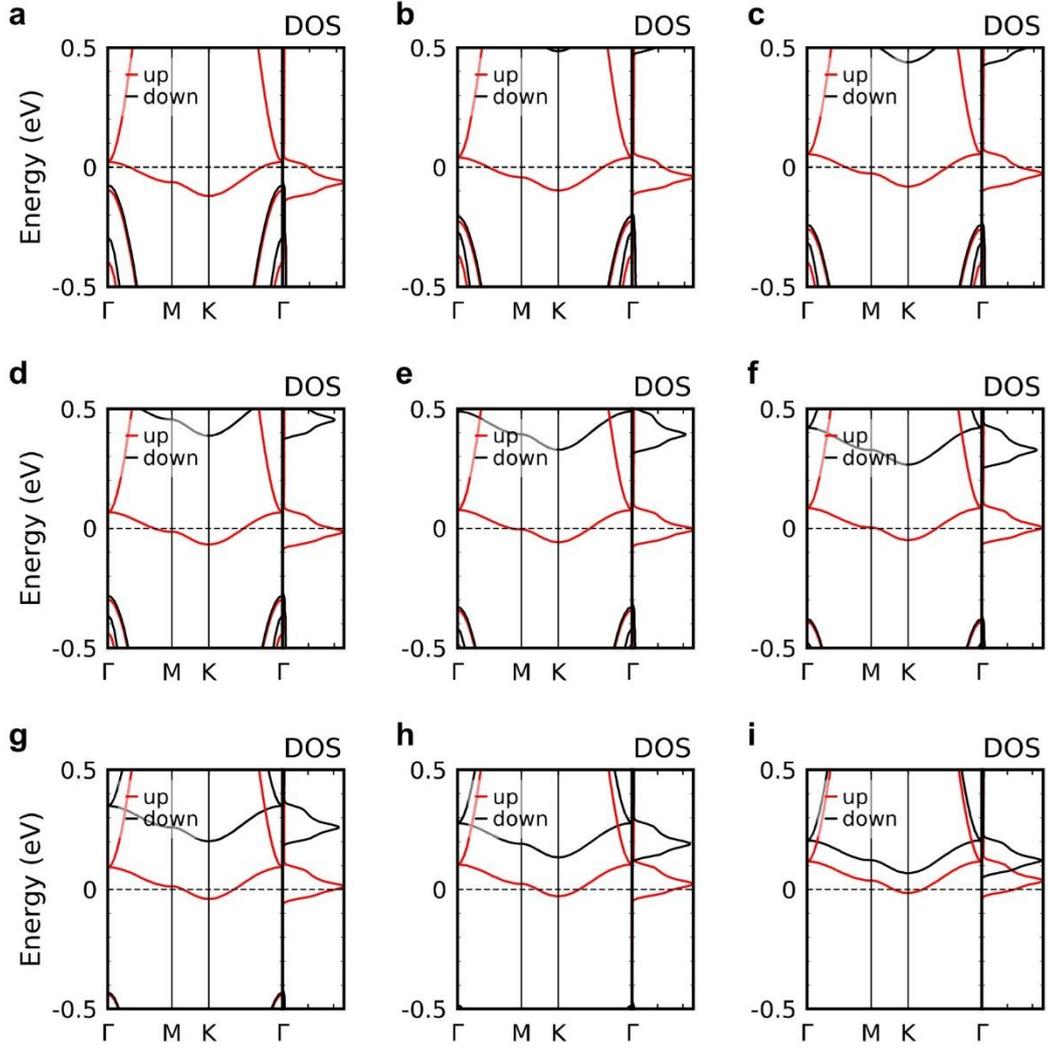

**Supplementary Figure 3.** The spin-polarized band structures of the $B_9$ *t*-Kagome lattice for various hole carrier concentrations under 18% tensile strain. The spin-up flat band is doped with the carrier concentrations of (a) 0.1, (b) 0.2, (c) 0.3, (d) 0.4, (e) 0.5, (f) 0.6, (g) 0.7, (h) 0.8, and (i) 0.9 hole/cell. The density of states (DOS) is given in arbitrary unit.

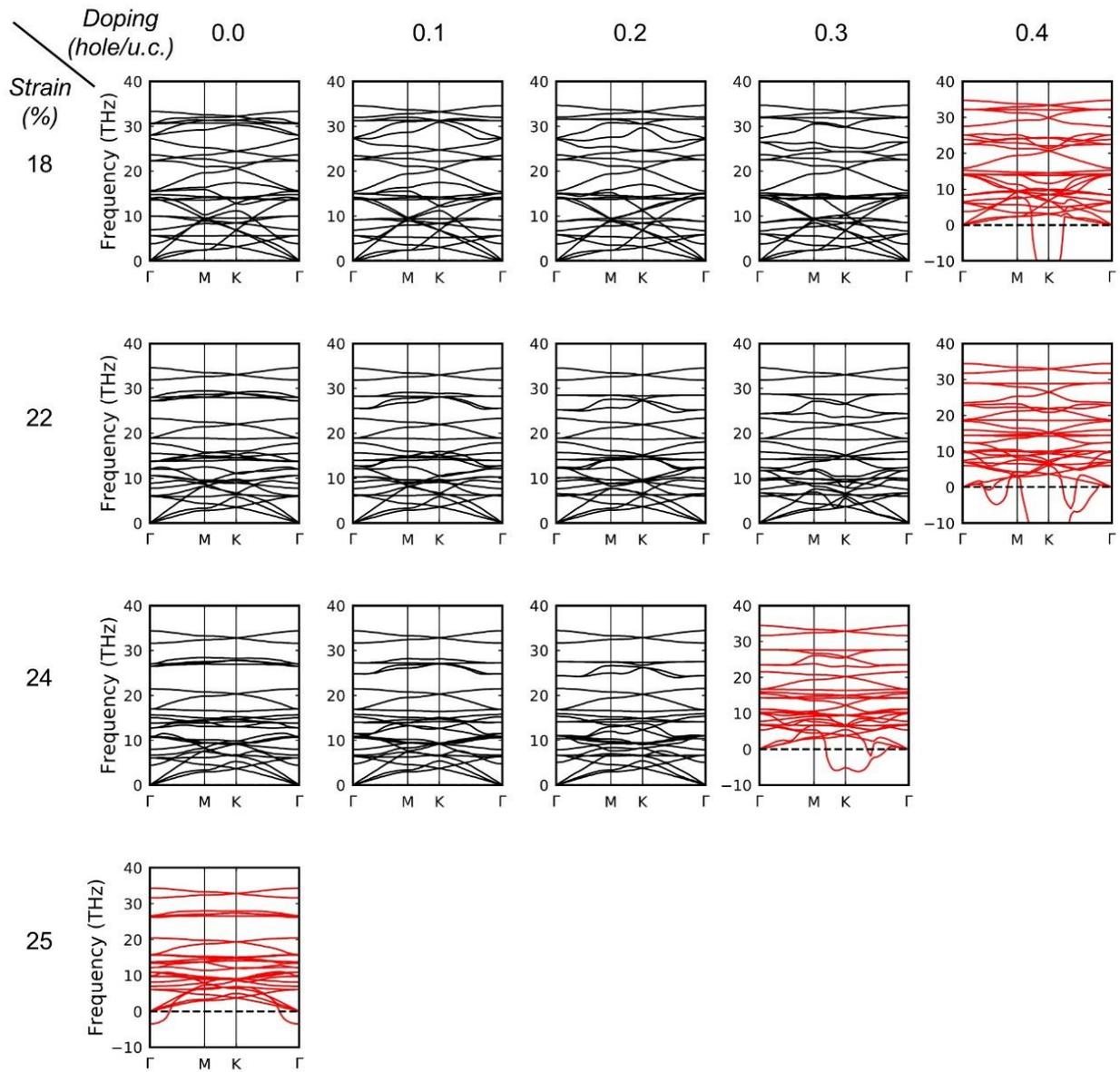

**Supplementary Figure 4.** The phonon dispersion curves of the B$_9$ *t*-Kagome lattice for various hole doping levels under different tensile strains. The red solid lines indicate the dynamical instability because of the imaginary phonon modes.

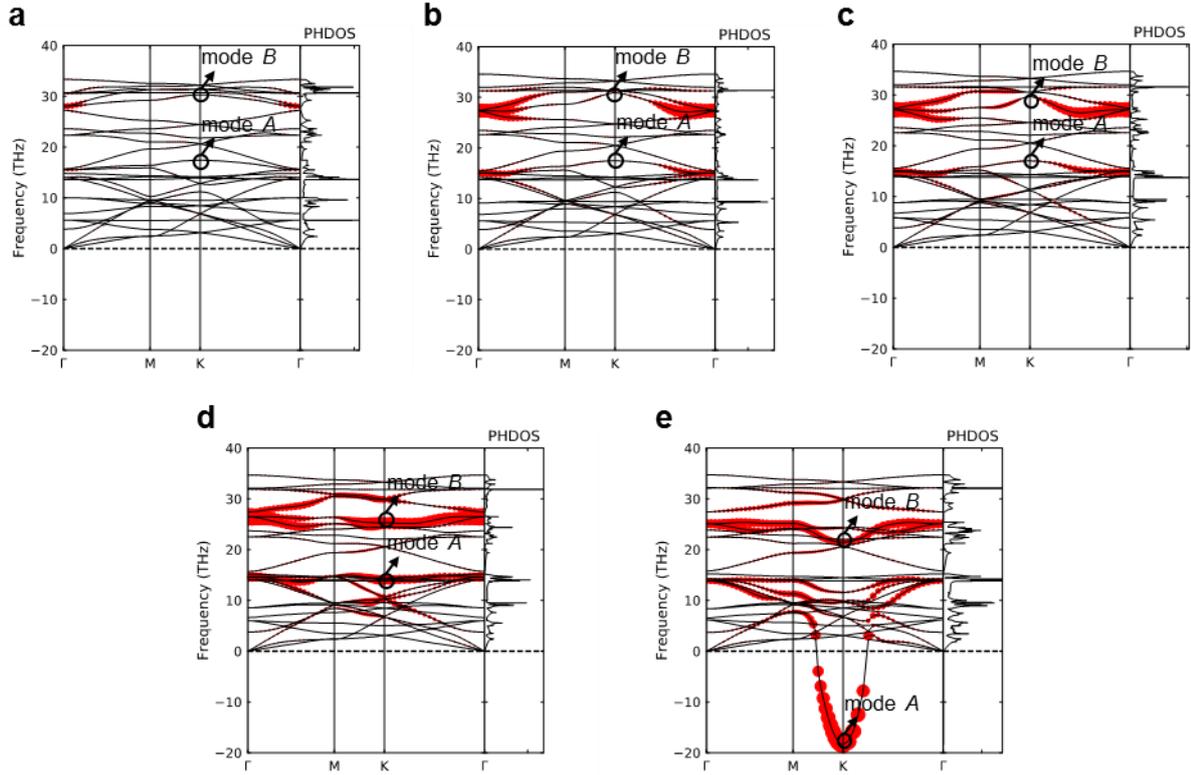

**Supplementary Figure 5.** The phonon dispersion curves and phonon densities of states (PHDOS in arbitrary unit) of the 18%-strained $B_9$ $t$-Kagome lattice for hole carrier concentrations of (a) 0, (b) 0.1, (c) 0.2, (d) 0.3, and (e) 0.4 hole/cell. The red circles represent the magnitude of the phonon linewidths from electron-phonon interactions. The localized phonon modes $A$ and $B$ have relatively large phonon linewidths. For the hole doping level of 0.4 hole/cell in (e), the imaginary mode $A$ occurs because of the strong electron-phonon interactions, indicating the dynamical instability.

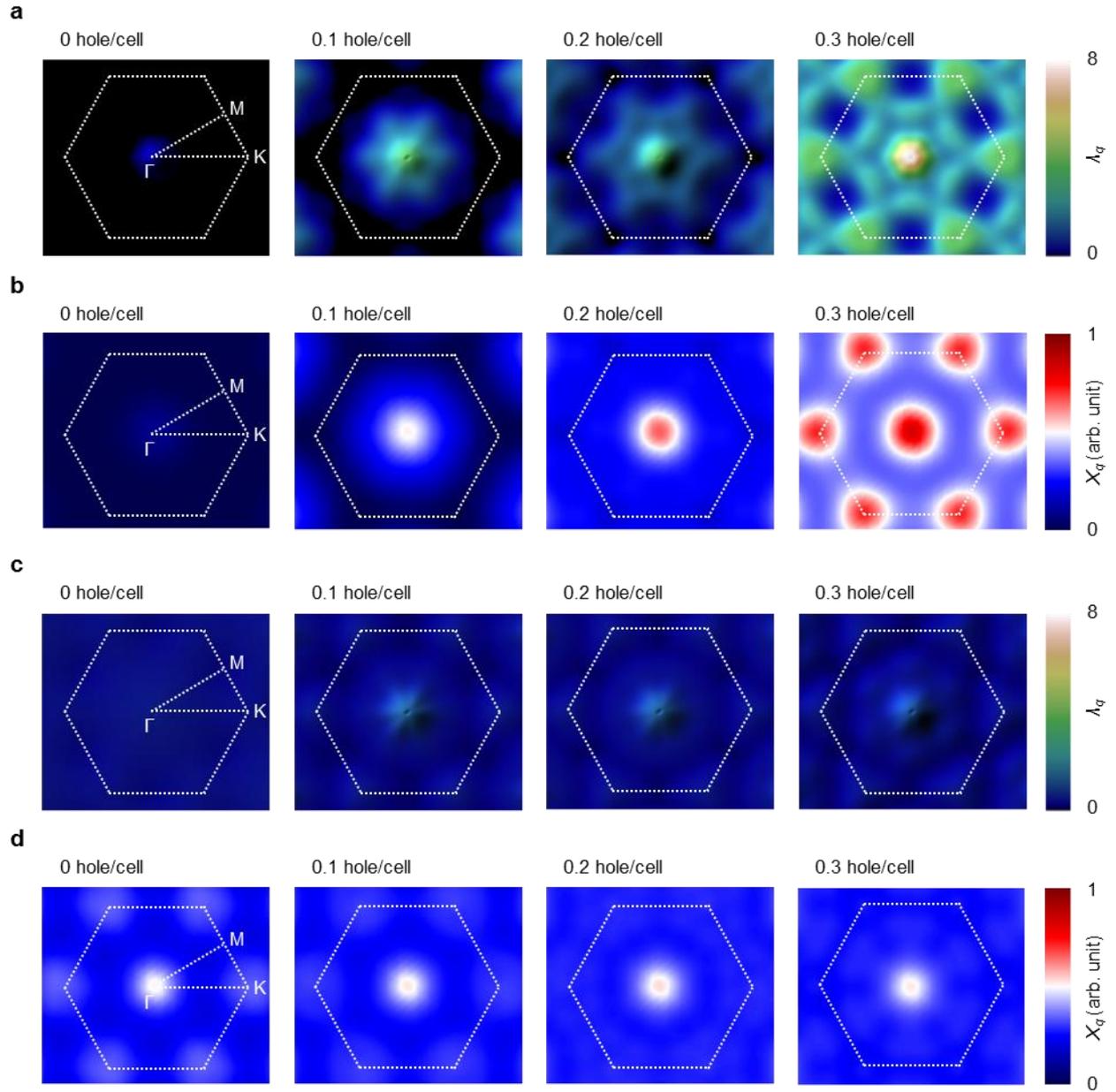

**Supplementary Figure 6.** The distribution of (a) momentum-dependent electron-phonon coupling ($\lambda_q$) and (b) nesting function ($\chi_q$) in hole-doped $B_9$ *t*-Kagome lattices under 18% strain through spin-polarized calculations. The results of spin-unpolarized calculations are shown in (c) and (d). The white dotted hexagons represent the 2D BZ.

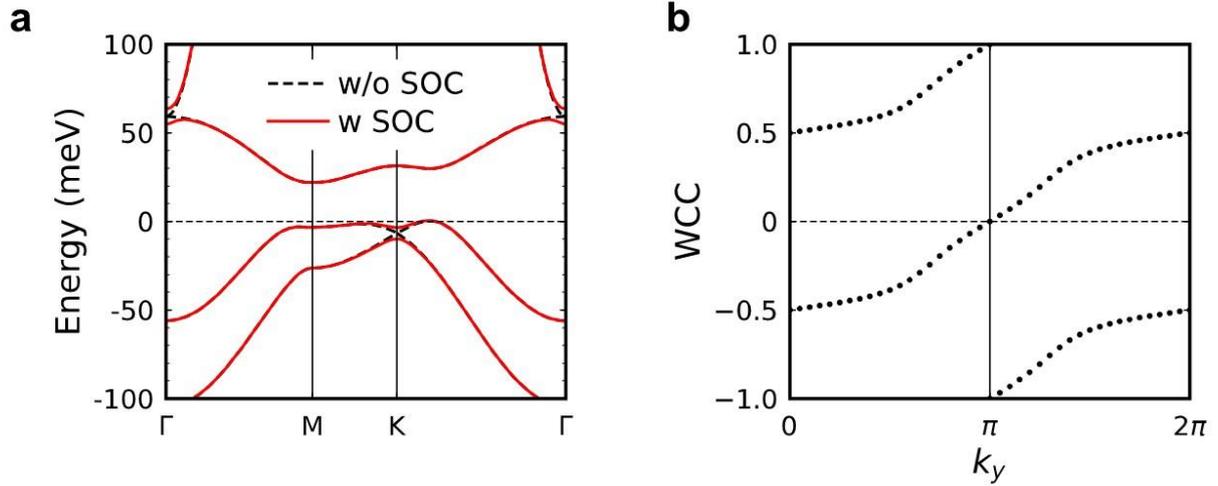

**Supplementary Figure 7.** (a) The effect of spin-orbit coupling (SOC) on the tight-binding band of the Wigner crystal formed at the hole carrier concentration of 0.666 hole/cell in the 18%-strained $B_9$ *t*-Kagome lattice. With the SOC, a band gap opens at the $\Gamma$ and $K$ points. (b) The evolution of the Wannier charge center (WCC) for the lowest occupied flat band in (a) along the $k_y$ axis. The winding number of the WCC is non-zero, indicating that the Chern number is 1 for the lowest occupied flat band. All calculations are performed using the pythTB package (http://physics.rutgers.edu/pythtb).

**Supplementary Table 1.** The hole carrier concentrations (in units of holes per cm$^2$) for various hole doping levels in the B$_9$ *t*-Kagome lattice under strain.

| | Doping (hole/cell) | Strain (%) | | | |
|---|---|---|---|---|---|
| | | 18 | 20 | 22 | 24 |
| | 0 | 0 | 0 | 0 | 0 |
| Hole carrier concentration (cm$^{-2}$) | 0.1 | 2.40×10$^{13}$ | 1.95×10$^{13}$ | 1.61×10$^{13}$ | 1.35×10$^{13}$ |
| | 0.2 | 4.80×10$^{13}$ | 3.89×10$^{13}$ | 3.22×10$^{13}$ | 2.70×10$^{13}$ |
| | 0.3 | 7.20×10$^{13}$ | 5.84×10$^{13}$ | 4.82×10$^{13}$ | 4.05×10$^{13}$ |
| | 0.4 | 9.61×10$^{13}$ | 7.78×10$^{13}$ | 6.43×10$^{13}$ | 5.40×10$^{13}$ |

**Supplementary Table 2.** The convergence test of the electron-phonon coupling constant (λ) and logarithmic averaged phonon frequency ($\omega_{\log}$) for various sets of the *k*-points in the $B_9$ *t*-Kagome lattices with different doping levels. For the 18%-, 20%-, and 22%-strained Kagome lattices, the 12×12 *k*-point grid is sufficient for the numerical convergence of the electron-phonon coupling constant. The numbers in parentheses are the results of spin-unpolarized calculations.

| k-points (n×n) | 18% strain | | | | | | | |
|---|---|---|---|---|---|---|---|---|
| | 0 hole/cell | | 0.1 hole/cell | | 0.2 hole/cell | | 0.3 hole/cell | |
| | λ | $\omega_{\log}$ (cm$^{-1}$) | λ | $\omega_{\log}$ (cm$^{-1}$) | λ | $\omega_{\log}$ (cm$^{-1}$) | λ | $\omega_{\log}$ (cm$^{-1}$) |
| n = 6 | 0.113 | 1015 | 0.755 | 687 | 1.097 | 669 | 2.014 | 607 |
| n = 8 | 0.108 | 840 | 0.789 | 594 | 1.079 | 620 | 2.078 | 561 |
| n = 12 | 0.114 | 628 | 0.79 | 565 | 1.051 | 654 | 2.013 | 585 |
| | (0.335) | (569) | (0.366) | (545) | (0.372) | (543) | (0.37) | (563) |
| | 20% strain | | | | | | | |
| | 0 hole/cell | | 0.1 hole/cell | | 0.2 hole/cell | | 0.3 hole/cell | |
| k-points (n×n) | λ | $\omega_{\log}$ (cm$^{-1}$) | λ | $\omega_{\log}$ (cm$^{-1}$) | λ | $\omega_{\log}$ (cm$^{-1}$) | λ | $\omega_{\log}$ (cm$^{-1}$) |
| n = 6 | 0.133 | 880 | 0.966 | 579 | 1.454 | 572 | 3.25 | 475 |
| n = 8 | 0.137 | 685 | 1.007 | 503 | 1.46 | 521 | 3.478 | 429 |
| n = 12 | 0.127 | 768 | 0.958 | 568 | 1.415 | 563 | 3.364 | 453 |
| | (0.404) | (521) | (0.416) | (507) | (0.422) | -506 | (0.423) | (518) |
| | 22% strain | | | | | | | |
| | 0 hole/cell | | 0.1 hole/cell | | 0.2 hole/cell | | 0.3 hole/cell | |
| k-points (n×n) | λ | $\omega_{\log}$ (cm$^{-1}$) | λ | $\omega_{\log}$ (cm$^{-1}$) | λ | $\omega_{\log}$ (cm$^{-1}$) | λ | $\omega_{\log}$ (cm$^{-1}$) |
| n = 6 | 0.155 | 794 | 1.355 | 566 | 1.954 | 518 | 6.936 | 327 |
| n = 8 | 0.294 | 695 | 2.388 | 529 | 2.798 | 473 | 4.785 | 336 |
| n = 12 | 0.161 | 631 | 1.195 | 520 | 1.95 | 494 | 8.104 | 287 |
| n = 16 | 0.167 | 623 | 1.199 | 516 | 1.955 | 491 | 8.085 | 288 |
| | (0.458) | (488) | (0.416) | (507) | (0.479) | (478) | (0.481) | (486) |